Size dependent electric voltage-controlled magnetic anisotropy in multiferroic heterostructures: Interface-charge and strain co-mediated magnetoelectric coupling

Jia-Mian Hu,[1,2] Ce-Wen Nan,[1] and Long-Qing Chen[2]

[1]*Department of Materials Science and Engineering, and State Key Lab of New Ceramics and Fine Processing, Tsinghua University, Beijing 100084, China*

[2]*Department of Materials Science and Engineering, Pennsylvania State University, University Park, Pennsylvania 16802-5005*

**Abstract**

We present a phenomenological scheme to study the size-dependent electric voltage-controlled magnetic anisotropy in ferromagnetic (FM)/ferroelectric (FE) heterostructures. The FM layers are either metallic Fe(001), Ni(001), Co(0001), or half-metallic (La, Sr)MnO$_3$ films. Two magnetoelectric mechanisms, i.e., interface-charge and strain-mediated couplings, are considered. We show that the interface-charge mediated coupling is the main mechanism for the magnetoelectric coupling when the FM film thickness is below a certain transition thickness $d_{\text{tr}}$ while the strain-mediated coupling dominates above $d_{\text{tr}}$.

Artificial multiferroic heterostructures of ferroelectric (FE) and ferromagnetic (FM) layers are of increasing interest due to the coupling between the magnetic and electric polarizations [1]. Of particular interest in the multiferroic heterostructures, electric voltage, rather than the usual current or magnetic field, can be directly used to control the magnetic anisotropy or magnetization direction via magnetoelectric (ME) coupling [1-3], which offers promising applications for novel spintronic or ME devices with much lower power consumption and higher speed. Examples include voltage-driven magnetic random access memories [4,5], logic circuits [6], and microwave devices [7]. Much effort has been devoted to achieve robust *room-temperature* ME coupling by virtue of such FE/FM heterostructures either through a strain-induced ME effect across an interface [2,8-10], or interface-charges driven ME effect [11,12], or magnetic exchange bias [13–16]. As demonstrated recently in the multiferroic heterostructures, a remarkable electric-voltage control of magnetic behavior of the magnetic nanostructures at *room temperature* can be achieved by such a strain-induced ME coupling, i.e., an external voltage in the ferroelectric layer causing a strain change across the interface and then altering the magnetic anisotropy of the magnetic layer via the magnetostriction. For example, a butterfly-shaped magnetization (*M*)–electric field (*E*) loop at room temperature has recently been observed in a multiferroic heterostructure with a $La_{0.7}Sr_{0.3}MnO_3$ (LSMO, 20-50 nm) thin film grown on $Pb(Mg_{1/3}Nb_{2/3})_{0.72}Ti_{0.28}O_3$ (PMN-PT), which tracks the butterfly-shaped strain–*E* loop of PMN-PT, demonstrating a strain-induced ME coupling across the LSMO/PMN-PT interface [8]. However, in a similar heterostructure of LSMO (4 nm)/$PbZr_{0.2}Ti_{0.8}O_3$ (PZT, 250 nm), a totally different square-shaped *M-E* hysteric loop has been observed at 100 K, illustrating interface-charge driven ME coupling [11]. This discrepancy raises important questions: why do the two heterostructures behave differently, or do these two different ME coupling mechanisms operate independently? In this work, we demonstrate these coupling mechanisms could coexist and tend to interact with each other at the interfaces. Specifically, the interface-charge mediated ME coupling exerts major influences for ultrathin FM films while the strain-mediated ME coupling operates at larger thickness, leading to a size dependent electric-voltage control of magnetic anisotropy.

In this Letter we present a phenomenological approach to investigate such size effect of the ME coupling in the multiferroic FM/FE heterostructure, where the influences of two mechanisms for the ME coupling, i.e., the interface-charge and strain-mediated coupling, are addressed. For illustration, we consider different FM films including either metallic Fe(001), Ni(001), Co(0001), or half-metallic (001)-oriented LSMO films, grown on a FE layer such as $BaTiO_3$(001). The results show that there is a transition thickness $d_{tr}$ for the FM films, i.e., in the FM/FE heterostructures with a thin FM film below $d_{tr}$, the interface-charge mediated coupling plays a major part, while the strain-mediated ME coupling predominates when the FM film thickness is larger than $d_{tr}$.

Consider a multiferroic structure with a FM thin film grown on a FE layer, and an electric voltage $V$ applied longitudinally across the FE layer. Then the total magnetic anisotropy energy $F_{tot}(V)$ of the FM film in a single-domain state is [5,10]

$$F_{tot}(V) = F_{mc} + F_{shape} + F_{me}(V) + F_S(V), \qquad (1)$$

where $F_{mc}$ is the magnetocrystalline anisotropy, $F_{shape}$ the shape anisotropy, $F_{me}$ the magnetoelastic anisotropy, and $F_S$ surface anisotropy, The strain-induced coupling across the interface and the interface-charge driven coupling are mainly related to the magnetoelastic anisotropy $F_{me}$ and the surface anisotropy $F_S$, respectively, where $F_S$ can be expressed as [17, 18],

$$F_{surf} = -\frac{2K_s + \Delta K_s(V)}{d} m_3^2, \qquad (2)$$

here $m_3$ refers to the direction cosine; $K_s$ and $\Delta K_s(V)$ denote the surface anisotropy energy and its change under external electric voltage $V$; and $d$ is the thickness of the FM film. Any changes in the interface charges would alter the surface anisotropy, and hence the magnetization state.

For simplicity, the effective magnetic anisotropy field $H_{eff}$ [7,19] is used to investigate the voltage-controlled magnetic anisotropy ($H_{eff}$ usually shares similar variation trend with the magnetic coercive field $H_c$ [20]). Thus the out-of-plane effective anisotropy field, i.e., $H_{eff}^{OP}$, can be determined by

$$H_{eff}^{OP} = -\frac{1}{\mu_0 M_s} \frac{\partial F_{tot}}{\partial m_3}\bigg|_{m_3=1}, \qquad (3)$$

where $\mu_0$ and $M_s$ are the vacuum permeability and the saturation magnetization, respectively. $H_{eff}^{OP}$ can be experimentally obtained from an out-of-plane magnetic hysteresis loop [21]. An out-of-plane magnetic easy axis (or spontaneous magnetization) is preferred for $H_{eff}^{OP} > 0$, and a change in the sign of $H_{eff}^{OP}$ from positive to negative would indicate an easy axis reorientation [10] from an out-of-plane to an in-plane direction, or vice versa. Such a reorientation depending on the film thickness has been reported in FM thin films [22-24]. For example, in a Ni/Cu(001) heterostructure [22], the easy axis of the magnetization switched abruptly from initial in-plane to out-of-plane at a critical thickness $d_{cr}$ of about 10.5 ML (~1.6 nm). The critical thickness $d_{cr}$ for such an easy axis reorientation can be estimated from $H_{eff}^{OP} = 0$, i.e., $d_{cr} = 2K_s \left/ \left[ \frac{1}{2}\mu_0 M_s^2 - K_1 - B_1(1 + 2c_{12}/c_{11})\varepsilon_0 \right] \right.$. By using the known material parameters [25], and the residual strain $\varepsilon_0$ of 2.5% arising from the in-plane lattice mismatch between Ni film and Cu(001) substrate [22], one can obtain $d_{cr}$ of about 1.73 nm for the Ni film, well consistent with the experimental value (~1.6 nm), demonstrating that this effective anisotropy field approach is valid.

Now let us return to the change in the magnetic anisotropy under the application of longitudinal electric voltages to the bilayer structure, i.e., $\Delta H_{eff}^{OP}$ $[= H_{eff}^{OP}(V)/H_{eff}^{OP}(0) - 1]$, which can be obtained as,

$$\Delta H_{eff}^{OP} = \frac{2\left[ B_1\left(1 + \frac{2c_{12}}{c_{11}}\right)\varepsilon_p(V) + \frac{\Delta K_s(V)}{d} \right]}{M_s} \bigg/ H_{eff}^{OP} \tag{4a}$$

for cubic (001) FM films, and,

$$\Delta H_{eff}^{OP} = \frac{2\left[ \left(B_1 + 2B_3 - \frac{2B_2 c_{13}}{c_{33}}\right)\varepsilon_p(V) + \frac{\Delta K_s(V)}{d} \right]}{M_s} \bigg/ H_{eff}^{OP} \tag{4b}$$

for hexagonal (0001) films. Here $\varepsilon_p(V)$ denotes the piezostrain under external voltage. The two terms in the square brackets on the right hand side of Eq. (4) describe the contributions from the strain and the interface-charge mediated ME coupling. Thus these two mechanisms coexist in the FM/FE bilayer structure and compete with each other.

For illustration, the calculations are performed for the FM films by using the known material parameters [25]; and a common BaTiO$_3$ (BTO) film with a thickness of about 100 nm grown on (001) SrTiO$_3$ substrate [26] is chosen as the FE layer, with SrRuO$_3$ as a bottom electrode. The butterfly-shaped curve of piezostrain $\varepsilon_p(V)$ shown in Fig. 1(a) was measured in the BTO film using a piezoelectric force microscope. For the metallic Fe(001) films, the hysteresis-like change in $\Delta K_s(V)$ [Fig. 1(b)] is directly obtained from the voltage-controlled magnetic anisotropy in ultrathin Fe atomic layers [17]. The same $\Delta K_s(V)$ is used for the Ni(001) and Co(0001) films for simplicity, since they were reported to exhibit similar surface ME coupling coefficients to the Fe(001) films [18] that also depend on the spin-polarized interface screening charges.

As the first example, the electric voltage-induced changes in $H_{eff}^{OP}$ for the (001) Fe/BTO thin-film heterostructure are shown in Fig. 2(a). It shows distinct size-dependent characteristics of the voltage-controlled out-of-plane effective anisotropy field, i.e., $\Delta H_{eff}^{OP}$, demonstrating the coexistence of both the interface-charge and strain mediated ME coupling in the heterostructure. A transition thickness $d_{tr}$ for the two interacting ME coupling mechanisms can be estimated to be about 0.5 nm (about one-unit cell thickness) as the contributions from the two mechanisms become equal from Eq. (4). Thus, when the Fe film thickness is smaller than $d_{tr}$, the $\Delta H_{eff}^{OP}$-voltage curve tends to mimic the voltage-induced surface anisotropy change behavior, i.e., a hysteresis-like loop [see Fig. 2(a)], indicating that the interface-charge mediated ME coupling could play a major part. However, the $\Delta H_{eff}^{OP}$-voltage loops become butterfly-shaped as the film thickness exceeds $d_{tr}$, presenting a dominant strain-mediated ME effect. An external voltage leads to large $\Delta H_{eff}^{OP}$ changes in Fe films with smaller thickness (i.e., below $d_{tr}$), as compared to those with thickness larger than $d_{tr}$. Such larger $\Delta H_{eff}^{OP}$ can in principle allow a more dramatic voltage-induced magnetic anisotropy change based on a dominative interface-charge ME coupling in the reduced thickness scale. Moreover, the maximal value of $\Delta H_{eff}^{OP}$ emerges at $d$=0.3 nm to be about 20% under the action of negative voltages, which could be attributed to the enhanced sensitivity of $\Delta H_{eff}^{OP}$ to both the external voltage and the residual strain when approaching the transition thickness $d_{tr}$, as illustrated in Fig. 2(b). Furthermore, it can be seen that the $\Delta H_{eff}^{OP}$-voltage loop reverses as the film thickness increases from 0.3 nm to 1 nm, owing to a sign change of $H_{eff}^{OP}$ at a critical thickness $d_{cr}$ of about 0.39 nm (not shown here) where the magnetic easy axis of the Fe film switches from an out-of-plane to an in-plane direction.

In comparison with the Fe(001) film, the Ni(001) film presents different behavior, as shown in Figure 3(a) for $\Delta H_{eff}^{OP}$ in the (001) Ni/BTO bilayer structure. A butterfly-shaped $\Delta H_{eff}^{OP}$-voltage curve is clearly shown even when the film thickness is reduced to 0.15 nm [see the inset of Fig. 3(a)], demonstrating the dominant influence of strain-mediated ME coupling in the (001) Ni/BTO heterostructure. Similarly to the (001) Fe/BTO case, the $\Delta H_{eff}^{OP}$-voltage curves reverse as the film thickness $d$ exceeds the critical $d_{cr}$ of about 1.76 nm (at $\varepsilon_0$=2.5% [22]) and exhibit enhanced $\Delta H_{eff}^{OP}$ in the vicinity of $d_{cr}$ where $H_{eff}^{OP}$ changes significantly. The (0001) Co/BTO structure presents also quite similar behavior [Fig. 3(b)], i.e., butterfly-shaped $\Delta H_{eff}^{OP}$-voltage behavior, demonstrating the dominant influence of strain-mediated ME coupling. Therefore, the interface-charge mediated ME coupling may not exist, but the strain-mediated ME coupling is always present in Ni(001) and Co(0001) films. Meanwhile, the strain-induced $\Delta H_{eff}^{OP}$ in these two cases are much larger than that in the (001) Fe/BTO structure due to their larger magnetoelastic coupling coefficients [25]. Moreover, it should be noted that these butterfly-shaped $\Delta H_{eff}^{OP}$-voltage loops in the (001) Ni/BTO and (0001) Co/BTO bilayers exhibit opposite trends. This is due to the opposite signs of $H_{eff}^{OP}$ and different $K_s$ [25] in the Ni(001) and Co(0001) films.

Now turn to an even more interesting FM/FE heterostructure consisting of FM half-metal like LSMO. LSMO is used due to its high sensitivity of a strongly correlated magnetic state to the charge carriers [11]. In half-metals, the screening interface-charges are usually 100% spin-polarized, which in principle allows stronger ME coupling as compared to the partial spontaneous spin polarization in FM metals such as Fe, Ni, and Co [27]. Figure 4(a) shows an electric-voltage dependent $\Delta K_s$ in the LSMO(001) film [27]. The sharp changes of $\Delta K_s$ in the vicinity of the FE coercive field can be related to the two distinct states for the spin-polarized interface charges resulting from the FE polarization reversal, i.e., the accumulation and the depletion state, respectively, as observed in a recent experiment [11]. Similarly to the case of the (001) Fe/BTO structure, the (001) LSMO/BTO heterostructure presents either hysteresis-like or butterfly-shaped $\Delta H_{eff}^{OP}$-voltage loops at room temperature [Fig. 4(b)], depending on the thickness of the LSMO films, exhibiting the interface-charge and strain co-mediated ME coupling. However, the room-temperature transition thickness $d_{tr}$ in the (001) LSMO/BTO case is significantly larger than that in the (001) Fe/BTO case, i.e., about 4.2 nm, owing to an enhanced ME coupling and thus the larger voltage-induced surface anisotropy change $\Delta K_s$ as discussed above. Furthermore, it can be seen that the change in $\Delta H_{eff}^{OP}$ is greater in LSMO films with thickness above $d_{tr}$ than that in LSMO thin films, indicating the voltage-controlled magnetic anisotropy change may be more significant in the thick LSMO(001) films where the strain-mediated ME coupling dominates, as observed in recent experiments [8,11]. The $\Delta H_{eff}^{OP}$-voltage curves of the LSMO/BTO structure at low temperature present a similar size-dependent behavior [Fig. 4(c)] to its room temperature case. However, it can be expected that the interface-charge mediated ME effect would become more remarkable at low temperature in comparison with the suppressed strain-mediated ME coupling due to reduced piezoelectric strains, which can further lead to a larger transition thickness $d_{tr}$, demonstrating a major influence of the interface-charge mediated ME coupling within a wider thickness range.

In conclusion, a simple phenomenological model is able to describe the size dependent electric voltage-controlled magnetic anisotropy in multiferroic heterostructures. The interface-charge and strain mediated coupling coexist and interact with each other in bilayer structures. A transition thickness $d_{tr}$ is defined to describe the competition between these two coupling mechanisms, below which the influence of the interface-charge mediated ME coupling would outweigh that of the strain-mediated ME coupling. The calculations show that interface-charge and strain co-mediated ME coupling can be clearly observed in the (001) Fe/BTO and the (001) LSMO/BTO structures. In particular, LSMO(001) films exhibit large transition thicknesses $d_{tr}$, indicating a more remarkable interface-charge mediated ME effect. While in the Ni(001) and the Co(0001) films, the interface-charge mediated ME coupling may not exist and the strain-mediated ME coupling is always dominant.

**Acknowledgements**

This work was supported by the NSF of China (Grant Nos. 50832003 and 50921061) and the National Basic Research Program of China (Grant No. 2009CB623303); and NSF under the grant number DMR-1006541.

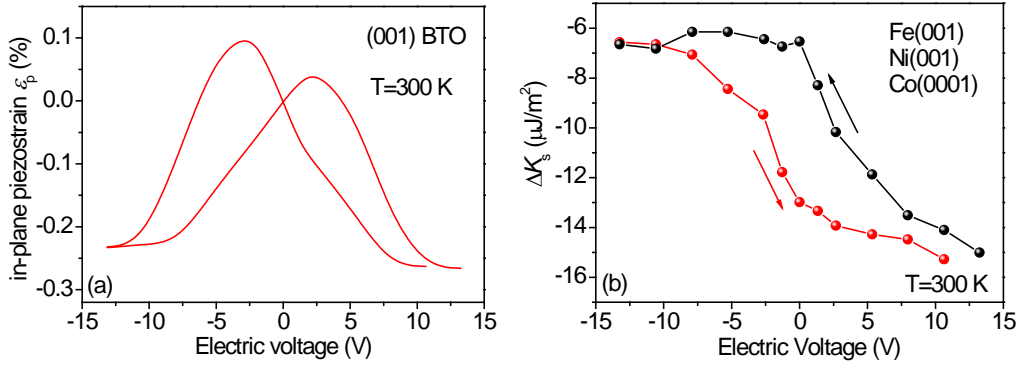

FIG. 1. Electric voltage dependence of (a) the in-plane piezostrain $\varepsilon_p$ generated in the (001) BaTiO$_3$ (BTO) film and (b) the surface anisotropy energy change $\Delta K_s$ in the Fe(001), Ni(001) and Co(0001) films.

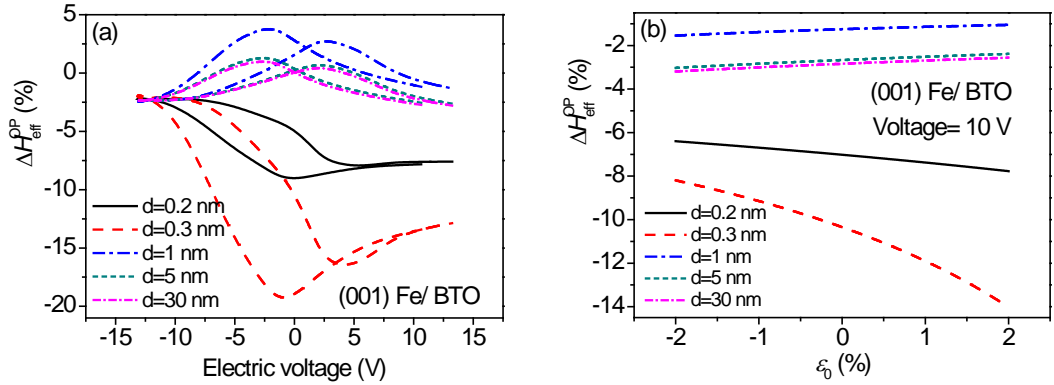

FIG. 2. (a) Electric voltage-induced change of the $H_{eff}^{OP}$, i.e., $\Delta H_{eff}^{OP}$, in the (001) Fe/BTO bilayers with various thicknesses $d$ of the Fe(001) thin films. (b) Variation trends of $\Delta H_{eff}^{OP}$ as a function of the residual strain $\varepsilon_0$ in the (001) Fe/BTO bilayers at an applied voltage of 10 V.

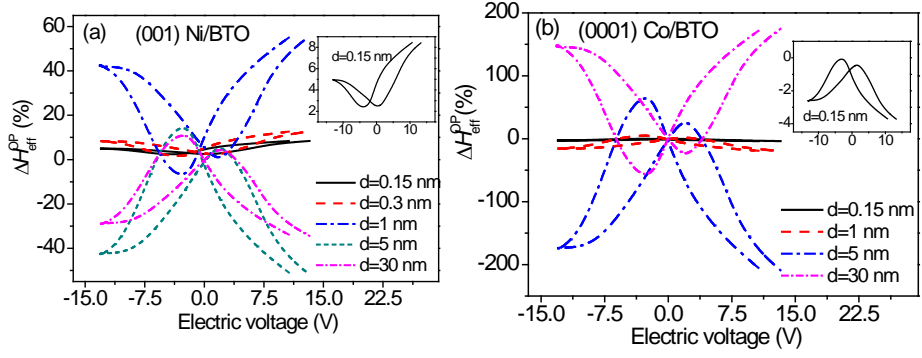

FIG. 3. Variations of the out-of-plane effective anisotropy field, i.e., $\Delta H_{eff}^{OP}$, with external electric voltages in (a) the (001) Ni/BTO and (b) the (0001) Co/BTO bilayers with different thicknesses $d$ of the Ni or Co thin films.

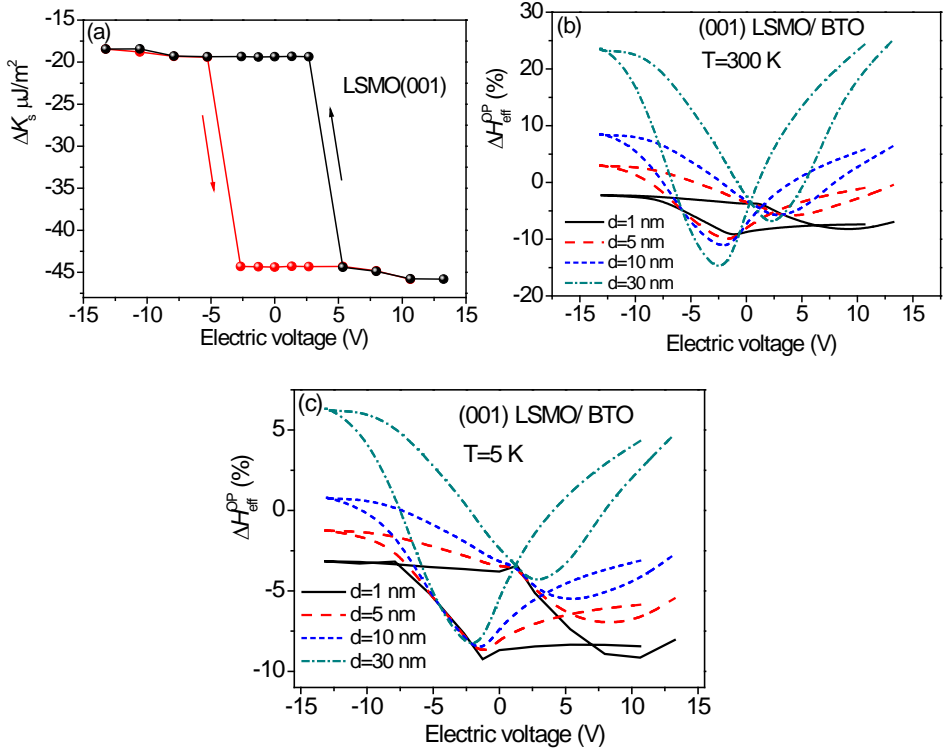

FIG. 4. (a) Electric voltage dependence of the surface anisotropy energy change $\Delta K_s$ in the (001)-oriented $La_{0.88}Sr_{0.1}MnO_3$ (LSMO) thin film. Electric voltage-induced change in the $H_{eff}^{OP}$, i.e., $\Delta H_{eff}^{OP}$, in the (001) LSMO/BTO bilayers with different thicknesses $d$ of the LSMO thin films, at (b) T=300 K and (c) T=5 K.